\documentclass{article}
\usepackage{amsmath, amssymb, amscd}
\title{Kolmogorov's Algorithmic Mutual Information Is Equivalent to Bayes' Law}
\author{Fouad B. Chedid \footnote{A'Sharqiyah University, Sultanate of Oman, Email: f.chedid@asu.edu.om.}}
\date{}
\begin{document}
\maketitle
\begin{abstract}
Given two events $A$ and $B$, Bayes' law is based on the argument that the probability of $A$ given $B$ is proportional to the probability of $B$ given $A$. When probabilities are interpreted in the Bayesian sense, Bayes' law constitutes a learning algorithm which shows how one can learn from a new observation to improve their belief in a theory that is consistent with that observation. Kolmogorov's notion of algorithmic information, which is based on the theory of algorithms, proposes an objective measure of the amount of information in a finite string about itself and concludes that for any two finite strings $x$ and $y$, the amount of information in $x$ about $y$ is almost equal to the amount of information in $y$ about $x$. We view this conclusion of Kolmogorov as the algorithmic information version of Bayes' law. This can be easily demonstrated if one considers the work of Levin on prefix Kolmogorov complexity and then expresses the amount of Kolmogorov mutual information between two finite strings using Solomonoff's a priori probability.
\end{abstract}

\section{Bayes' Law}
Bayes' law, also known as Bayes' rule or Bayes' theorem, proposes a way to make use of available data that may affect the likelihood of an event to better assess the probability of occurence of that event. This law was proposed by the Reverand Thomas Bayes, born in the early 1700s, but this work of Bayes wasn't made public during his lifetime, and it wasn't until after his death in 1761 that his student Richard Price communicated Bayes' work to John Canton in 1763 suggesting that ``a communication of it to the Royal Society cannot be improper"\cite{price}.

In the Bayesian interpretation, where a probabilty is a measure of our degree of belief in something which is different from the frequentist interpretation of probability, given a phenomenon $P$ and a proposed theory $T$ for $P$, Bayes' law provides a tool that quantifies the validity of $T$ as supported by our initial belief (a subjective measure) in $T$ and the observation of some evidence $E$ about $P$. As an equation, Bayes' law is stated as follows. 
\begin{equation}  
\label {bayes}
P(T|E) = \frac {P(T)}{P(E)}.P(E|T)
\end{equation}
We note that the above mathematical form of Bayes' law is due to the French mathematician Laplace who came to a similar conclusion in 1774. Following Laplace's logic, equation \ref{bayes} argues that the probability of a theory given an evidence is proportional to the probability of that evidence given that theory. In equation \ref{bayes}, $P(T)$ is the initial probability of $T$, which is our initial belief in $T$ prior to observing any evidences (aka a priori probability or the prior), $P(T|E)$ is the probability of $T$ given $E$, which is the probability of $T$ after accounting for the evidence $E$ (aka a posteriori probability or the posterior), and $P(E)$ is the probability of $E$ given all possible theories for $P$. Thus, Bayes' law allows us to update our initial belief in $T$ in a way that accounts for the evidence $E$. 

Clearly, Bayes' law constitutes a learning algorithm and is probably one of the earliest demonstration of a data-driven approach to learning. We view the difference between the a priori probability and the posteriori probability in Bayes' law as some manifestation of the amount of information in $T$ about $E$. Informally put, we have
\begin{equation}
P(T) - P(T|E) = \text { A manifestation of the amount of information in }T\text{ about }E.
\end{equation}

In \cite{price}, Price writes about the problem solved by Bayes' law that ``Every judicious person will be sensible that the problem now mentioned is necessary to be sovled in order to a sure foundation for all our reasonings concerning past facts, and that is likely to be hereafter". In fact, Bayes' law is considered one of the most fundamental applications of probability theory and has been compared to the Pythagorean theorem in geometry \cite{jeffreys}.

\section{Kolmogorov's Algorithmic Information}
Kolmogorov Complexity (aka Algorithmic Information Theory) was developed independently by Ray Solomonoff \cite{solo}, Andrey Kolmogorov \cite{kolm}, and Gregory Chaitin \cite{chaitin} in 1964, 1965, and 1969, respectively. At the core of this theory is the notion of a Universal Turing Machine of Alain Turing \cite{turing36}, which follows from the fact that a Turing machine is capable of simulating any other Turing machine. 
\subsection{Why Universality Matters?}
Picture yourself walking into a bookstore. There, you would find people of all walks of life browsing books either for fun or for learning enough about the information content of a book so that they can decide whether the content is worth the price. In this context, the quality of a book is not an intrinsic property of the book itself, but rather depends on the background and taste of the reader, which naturally would differ from one reader to another.  Now, obtaining an objective measure of the quality of any book requires that we have a universal reader who is a highest authority in all subjects. A book review provided by a universal reader could then be described as an intrinsic property of the book itself.  

All three inventors (Kolmogorov, Solomonoff, and Chaitin) used the concept of a Universal Turing Machine (UTM) to propose new ideas with built-in universal qualities. In particular, Kolmogorov's notion of algorithmic information relies on the existence of a universal decompressor (universal Turing machine) to propose a new definition of information that deviates from the notion of information being tied to a random variable (as discussed in Shannon's information theory) and makes information tied to an individual string, free of any probability distributions. Similarly, Solomonoff's notion of algorithmic probability is the halting probability of a Universal Probabilistic Turing Machine (UPTM) that takes no inputs, or equivalently, the halting probability of a UTM that takes an infinite random string as an input.
\subsection{Kolmogorov Complexity} 
Kolmogorov used the theory of algorithms of Turing to redefine randomness as incompressibility (or equivalently, lack of regularities) and to propose that the random or incompressible content of a finite string represents the amount of uncertainty or information in it. Given a finite string $x$, Kolmogorov complexity is defined as our ability to capture the regular part of $x$ so that when given the random part $p$ of $x$, we would be able to reconstruct $x$ from $p$ (decompresses $p$ to regenerate $x$). We note that we are not interested in how $x$ got compressed down to $p$, we simply want to have an effective way to uncover $x$ from $p$\footnote{Kolmogorov complexity is about decompression, and not compression.}. 

Our discussion this far suggests that the Kolmogorov complexity of a finite binary string $x$ is defined relative to a particular decompression algorithm (a Turing machine) $M$ as the length of a smallest input $p$ that causes $M$, when it reads $p$, to generate $x$ and then halts. In a mathematical form, we have
\begin{equation}
\label{kolcomp}
C_M(x) = \min\limits_{\text {input }p}\{|p|: M(p)=x\}
\end{equation}
As such, Kolmogorov complexity is made relative to a particular algorithm or Turing machine, which can hardly mean anything (this is similar to the point we made in the previous section when we argued that the quality of a book cannot be determined by the taste and background of a particular reader). Here, the notion of a Universal Turing Machine (UTM) comes to the rescue. Knowing that a UTM $U$ can simulate any other Turing machine $M$, rewriting equation \ref{kolcomp} relative to $U$ gives a universal meaning to $C_U(x)$. This is true because we can easily show that for all other Turing machines $M$, we have 
$$ C_U(x) \le C_M(x) + c $$  
where $c$ is a constant that depends on $M$, but not on $x$. In particular, $c$ is about the length of the binary encoding of $M$. The input of $U$ would consist of the pair $(\text{bin}(M),p)$ so that $U$ would know how to simulate $M$ on the input $p$; that is, $U(\text{bin}(M),p)=M(p)$. If we next let $C_M(x)=|q|$ for some string $q$; that is, $q$ is a shortest input for $M$ to generate $x$, then $C_U(x) \le q + |\text{bin}(M)| + c^{\prime}$, where $c^{\prime}$ is the length of a self-delimiting binary encoding of the length of $\text{bin}(M)$, which $U$ uses to separate bin$(M)$ from $q$. This result is known as the Invariance Theorem and was discovered independently by Solomonoff and Kolmogorov. Moreover, given any other UTM $U^{\prime}$, since a UTM is just another Turing machine, we have
$$ C_U(x) \le C_{U{\prime}}(x) + c $$
$$ C_U{\prime}(x) \le C_U(x) + c $$
or equivalently, 
$$ | C_U(x) -  C_{U{\prime}}(x)| \le c $$
where $c$ is a constant that depends on $U$ and $U^{\prime}$, but not on $x$. Thus, it doesn't really matter which UTM we choose in our definition of Kolmogorov complexity as long as we accept to tolerate an additive constant error in the result, which can be large! A better statement is that it actually does matter which UTM we use in the definition of Kolmogorov complexity, but once we fix a reference UTM $U$, we will have a universal definition in the sense that the value of $C_U(x)$ may exceed the true value of the amount of information in $x$ by a constant term, but it is never less than it.  

\subsection{Kolmogorov Mutual Information}
First, we review the argument of Kolmogorov for calling $C(x)$ the amount of information in $x$ about itself. Kolmogorov introduced the notion of the conditional complexity of a string $x$ in presence of another string $y$ that is made available to the UTM $U$ for free. In particular, we have 
$$ C_U(x|y) = \min\limits_{\text {input }p}\{|p|: U(y,p) = x\}$$  
Here, we follow Kolmogorov's notation and place the auxilliary information $y$ before the input $p$. Next, Kolmogorov argued \cite{kolm} that since $C(x|y)\le C(x)$, it is fair to call the differene $C(x)-C(x|y)$ the amount of information in $y$ about $x$, to be denoted by
 $$I_U(y:x) = C_U(x) - C_U(x|y)$$ 
We note that this argument is similar to the argument we hinted to in Section 1 on Bayes' law, when we proposed $P(T)-P(T|E)$ as a manifestation of the amount of  information in $E$ about $T$.

We next ask what is $I(x:x)$? What is $C(x:x)$?\\
Clearly, $C(x|x) = \min\limits_{\text {input }p}\{|p|: M(x,p) = x\} = $ the constant length of the copy program $p$ that copies its input to its output. Thus, up to an additive constant, $C(x|x) = 0$ and $I(x:x)=C(x)$. For this reason, Kolmogorov suggested to call $C(x)$ the amount of information in $x$ about itself\footnote{We could call it self-information, similar to the notion of self-entropy.}. 

We mention that prior to this work of Kolmogorov, the notion of the information content of a string wasn't there (almost). For example, Shannon's work was about the minimum number of bits needed on the average to transmit a value taken by a random variable, as a syntactic unit independent of any semantic\footnote{In his 1948 paper, Shannon wrote `` ... semantic aspects of communication are irrelevant to the engineering problem ..''. }. Similarly, Chaitin was interested in studying the size of a shortest program capable of generating a given sequence of bits on a universal Turing machine, and Solomonoff was interesed in predicting the next value taken by a random variable following an unknown probability distribution. While this is true in general, the following definition of a possible information measure was first suggested by Wiener in 1948: ``The amount of information provided by a single message $m_i$, $I(m_i) = -\log_2 p_i$'', which is related to the number of bits needed to identify any of the messages which happen to occur with probability $p_i$. The comforting thing is that both Shannon and Kolmogorov notions of information agree that {\em information is about removal of uncertainty}. This agrees with the point of view suggested by Kolmogorov, though seen from an opposite end, that information is about the ability to uncover regularity. That is, the more regular a string is, the less information it has\footnote{Alternatively, the less uncertainty it contains.}, and vice versa. 
\section{Solomonoff's Algorithmic Probability and Its Relationship to Kolmogorov Complexity}
To understand Solomonoff's algorithmic probability, we first need to recall the notion of a probabilistic Turing Machine (PTM). A PTM is similar to a non-deterministic Turing machine with an added read-only tape, called the random tape, that is full of random bits. The machine can have two possible next moves in any configuration, and the choice is made based on the next bit read off the random tape (the assumption is that the two possible next moves in any configuration are equally likely). Let $q$ be the random sequence of bits read off the random tape of a PTM $M$ when it runs on input $p$. The halting probability of $M(p,q)$ is the product of the probabilities of the choices taken at each step of the computation, which is $2^{-|q|}$. The output of $M$ is a string $x=M(p,q)$, if $M$ accepts $p$. We note that $M$ may accept $p$ in one execution and reject it in another. We next consider the case when $M$ runs on the empty tape. The halting probability $P_M(x)$ of $M$ is the sum of halting probabilities of $M(\lambda,q)$ for all random strings $q$ which cause $M(\lambda,q)$ to output $x$ and halt. Solomonoiff called $P_M(x)$ the algorithmic probabilty of $x$ relative to $M$. We have
$$P_M(x) = \sum\limits_{\text {random }q: M(\lambda,q)=x} 2^{-|q|}$$
Given that we require $M$ to halt immediately after it outputs $x$, then all random $q$ which appear in the equation of $P(x)$ must be prefix free. By Kraft's inequality, we have $P(x)\le 1$.

Solomonoff's work on algorithmic probability assumes a deterministic Turing machine $M$ (not a probabilistic one) whose input consists of an infinite random binary string with equal probabilities for zero and one. Thus, we can write 
$$P_M(x) =\sum\limits_{\text{random }p: M(p)=x}2^{-|p|}$$
Solomonoff next used the notion of a Universal Turing Machine (UTM) to give his algorithmic probability a universal sense and called his algorithmic probability $P_U(x)$ relative to a UTM $U$ a universal a priori $m(x)$. Levin \cite{levin} showed that $m(x)$ is a universal lower semicomputable semimeasure in the sense that for any other probabilistic Turing machine $M$, $P_M(x) \le c.m(x)$, for all $x$, where $c$ is a constant that depends on $M$, but not on $x$. This result shows that $m(x)$ dominates (is superior to) any other lower semicomputable semimeasure $P_M(x)$ (= the halting probability distribution generated by a probabilistic Turing machine). 

We recall that Solomonoff's overall objective was to be able to predict the next sequence of bits in a string that is generated by a random source for which we know nothing about its governing probability disribution. His method uses Bayes' law where the unknown a priori probability gets replaced by his a priori probability $m(.)$.

We conclude this section by emphasizing the following observations:
\begin{enumerate}
\item Setting the halting probability of $M$ to $2^{-|p|}$ makes the argument that the less (more) randomness the string $x$ contains, the higher (lower) its algorithmic probability is.
\item The algorithmic probability accounts for all possible different random contents that allows the machine $M$ to recover $x$. Solomonoff writes in \cite{solo}: ``The assignment of high a priori probabilities to sequences with many descriptions corresponds to a feeling that if an occurrence has many possible causes, then it is more likely."
\item Solomonoff's logic in his algortihmic probability agrees with the logic of Epicurus, which states that ``If more than one theory is consistent with the data, keep them all."  which intrestingly expresses the opposite sentiment to Occam's razor adopted by Kolmogorov in his definition of $C_M(x)$, which considers only a shortest random content of $x$ that allows $M$ to recover $x$.
\item Strings of high (low) algorithmic probability correspond to strings of  low (high) Kolmogorov complexity.  
\end{enumerate}
\section{Algorithmic Mutual Information Is Equivalent to Bayes' Law}
Levin showed that Solomonoff's algorithmic probability is related to a special type of Kolmogorov complexity, named prefix Kolmogorov complexity (discovered independently by Levin and Chaitin), which requires programs and inputs  to be prefix-free \cite{bienvenu,shen}. It is known that for every Turing machine $M$, one can construct an equivalent prefix-free Turing machine $M^{\prime}$ such that for all inputs $p$, $M(p)=M^{\prime}(p)$. The prefix Kolmogorov complexity of $x$, denoted by $K_U(x)$, is the length of a shortest input that causes a fixed reference prefix-free UTM $U$ to print $x$ and then halts. In the rest of this section, we use the notation $K(x)$ for $K_U(x)$ dropping the subscript $U$.

An important result of Levin \cite{shen} shows that up to an additive constant, for all finite strings $x$,
$$ -\log m(x) \ge K(x)$$ 
Combining this result with two simpler results (each is expressed up to an additive constant), namely, $-\log m(x) \le C(x)$ and $C(x) \le K(x)$, we conclude that up to an additive constant,
$$ -\log m(x) = K(x)$$
Using prefix Kolmogorov complexity to express the amount of information in a string about another, we conclude that up to an additive constant, $$I(y:x) = I(x:y)$$
This is true because up to an additive constant, we have 
$$ K(x,y) = K(x) + K(y|x) = K(y) + K(x|y)$$
which implies that
$$ K(x) - K(x|y) = K(y) - K(y|x)$$
or equivalently
$$ I(y:x) = I(x:y)$$
We view this argument that up to an additive constant the amount of information in $y$ about $x$ is equal to the amount of information in $x$ about $y$ as the algorithmic information version of Bayes' law. In fact, one can easily uses this argument to derive Bayes' law for Solomonoff's a priori probability. We have
$$K(x)-K(x|y) = K(y)-K(y|x)$$
Replacing $K(x)$ by $-\log m(x)$, we have, up to an additive constant
$$ -\log m(x) + \log m(x|y) = -\log m(y) + \log  m(y|x)$$ 
Applying basic rules for logarithms gives 
$$ m(x|y) = \frac {m(x)}{m(y)}.m(y|x)$$

\end{document}